\RequirePackage{fixltx2e}
\documentclass[twocolumn,english,journal]{revtex4-1}
\usepackage[T1]{fontenc}
\usepackage[latin9]{inputenc}
\usepackage{color}
\usepackage{babel}
\usepackage{calc}
\usepackage{units}
\usepackage{amsmath}
\usepackage{amssymb}
\usepackage{graphicx}
\usepackage{esint}
\usepackage[unicode=true,
 bookmarks=true,bookmarksnumbered=true,bookmarksopen=false,
 breaklinks=false,pdfborder={0 0 0},pdfborderstyle={},backref=false,colorlinks=true]
 {hyperref}
\hypersetup{pdftitle={Ferromagnetic Resonance in metastable magnetic states for the analysis of the specific heat of low energy spin excitations},
 pdfauthor={Benjamin W. Zingsem}}

\makeatletter

\newcommand{\lyxdot}{.}

\makeatother

\begin{document}
\title{Specific Heat of Spin Excitations Measured by Ferromagnetic Resonance}
\author{Benjamin W. Zingsem}
\affiliation{Faculty of Physics and Center for Nanointegration (CENIDE), University
Duisburg-Essen, 47057 Duisburg, Germany}
\affiliation{Ernst Ruska Centre for Microscopy and Spectroscopy with Electrons
and Peter Grünberg Institute, Forschungszentrum Jülich GmbH, 52425,
Jülich, Germany}
\author{Michael Winklhofer}
\affiliation{School of Mathematics and Science, Carl-von-Ossietzky University Oldenburg,
26129 Oldenburg}
\author{Sabrina Masur}
\affiliation{Cavendish Laboratory, Department of Physics, JJ Thomson Avenue, Cambridge
CB3 0HE, United Kingdom}
\author{Paul Wendtland}
\affiliation{Faculty of Physics and Center for Nanointegration (CENIDE), University
Duisburg-Essen, 47057 Duisburg, Germany}
\author{Ruslan Salikhov}
\affiliation{Faculty of Physics and Center for Nanointegration (CENIDE), University
Duisburg-Essen, 47057 Duisburg, Germany}
\affiliation{Helmholtz-Zentrum Dresden-Rossendorf, Institute of Ion Beam Physics
and Materials Research, Bautzner Landstrasse 400, 01328 Dresden, Germany}
\author{Ralf Meckenstock}
\affiliation{Faculty of Physics and Center for Nanointegration (CENIDE), University
Duisburg-Essen, 47057 Duisburg, Germany}
\author{Michael Farle}
\affiliation{Faculty of Physics and Center for Nanointegration (CENIDE), University
Duisburg-Essen, 47057 Duisburg, Germany}
\affiliation{Kirensky Institute of Physics, Federal Research Center \textquotedbl Krasnoyarsk
Science Centre, Siberian Branch of the Russian Academy of Science}
\begin{abstract}
Using ferromagnetic-resonance spectroscopy (FMR), we investigate the
anisotropic properties of epitaxial $\unit[3]{nm}$Pt/$\unit[2]{nm}$Ag/$\unit[10]{nm}$Fe/$\unit[10]{nm}$Ag/GaAs(001)
films in fully saturated meta-stable states at temperatures ranging
from $\unit[70]{K}$ to $\unit[280]{K}$. By comparison to spin-wave
theory calculations, we identify the role of thermal fluctuation of
magnons in overcoming the energy barrier associated with these meta-stable
states. We show that the energy associated with the size of the barrier
that bounds the meta-stable regime is proportional to the heat stored
in the magnonic bath. Our findings offer the possibility to measure
the magnonic contribution to the heat capacity by FMR, independent
of other contributions at temperatures ranging from $\unit[0]{K}$
to ambient temperature and above. The only requirement being that
the selected sample exhibits magnetic anisotropy, here, magnetocrystalline
anisotropy.
\end{abstract}
\maketitle

\section{Introduction}

Knowledge of the magnetic contribution to the thermal properties of
physical systems is essential for understanding magnetocalorics\citep{Gutfleisch2011},
spintronics\citep{Bader2010,Chumak2015}, and magnonic\citep{Chumak2015,Serga2014}
applications. Also, in biomedical applications like hyperthermia \citep{Myrovali2016,LiebanaVinas2016},
it is essential to assess the coupling between the magnetic contributions
and the crystal system's thermal properties. However, measuring the
contribution of magnons to the heat capacity of magnetic materials
at temperatures of technical interest is challenging. One approach
is to suppress magnons by shifting the spin-wave dispersion to energies
higher than the thermal energy. The magnon contribution to the total
heat can then be separated by conventional calorimetry experiments
with and without an applied field. Doing so for a system like Yittrium
Iron Garnet (YIG), however, would already require fields as high as
$\unit[30]{T}$ \citep{Rezende2015} at a temperature of only $\unit[20]{K}$.
Hence, it is unlikely that such experiments can get close to ambient
temperature. Another option is to measure the thermal population of
the spin-wave dispersion using, for example, Brillouin light scattering
(BLS). Here we propose a new method for determining the magnonic heat
capacity using ferromagnetic resonance (FMR). Our method is based
on ferromagnetic resonance measurements at a fixed frequency, where
a magnetic field is applied in different directions near a hard axis
of the magnetic anisotropy. First, the magnet is fully saturated,
then the direction of the field is swept, such that this saturated
configuration becomes metastable all the while FMR is measured at
each angle step. We then observe the transition from this metastable
configuration to a stable configuration in the FMR signal. From these
unconventional FMR measurements, we have evaluated the Zeeman energy
in critical configurations of metastable states. We find that in the
observed temperature regime between $\unit[70]{K}$ and $\unit[280]{K}$
the temperature-dependent change of the critical Zeeman energy is
proportional to the magnonic heat capacity.

\section{Experimental procedure}

Ferromagnetic resonance measurements were conducted on an epitaxial
$\unit[3]{nm}$Pt/$\unit[2]{nm}$Ag/$\unit[10]{nm}$Fe/$\unit[10]{nm}$Ag/GaAs(001)
thin film in which the (001)-direction of the Fe layer points out
of the sample plane. For such systems the Helmholtz free energy density
is known to have the form
\begin{equation}
\begin{aligned}F\left(\vec{B},\vec{M}\right)=-\vec{M}\left(\theta,\phi\right)\cdot\vec{B}\left(\theta_{B},\phi_{B}\right)+\frac{1}{2}\mu_{0}M^{2}\cos^{2}\left(\theta\right)\\
+2K_{2}^{\bot}\sin^{2}\left(\theta\right)+\frac{1}{4}K_{4}\left(\sin^{2}\left(2\theta\right)+\sin^{4}\left(\theta\right)\sin^{2}\left(2\phi\right)\right)
\end{aligned}
\label{eq:Free-Energy-Density}
\end{equation}
including the Zeeman contribution $\vec{M}\cdot\vec{B}$ as discussed
in \citep{Farle1998}. The angles are given in spherical coordinates,
where $\theta$ is the polar out-of-plane, and $\phi$ the azimuthal
in-plane angle of the magnetization, $\theta_{B}$ and $\phi_{B}$
are the polar and azimuthal angle of the external magnetic field $\vec{B}$
respectively. The anisotropy parameter $K_{4}$ accounts for the cubic
magnetocrystalline anisotropy, $K_{2}^{\bot}$ for a uniaxial anisotropy.
The thin-film shape-anisotropy is described by the demagnetization
term $\frac{1}{2}\mu_{0}M^{2}\cos^{2}\left(\theta\right)$.

For the magnetization to be resonantly excited, $\vec{M}\left(\theta,\phi\right)$
must be a minimizer of this energy landscape. In this case, the magnetization
is in an equilibrium state and the resonance condition is described
by the curvature of the energy landscape Eq. \ref{eq:Free-Energy-Density}
\citep{Zingsem2017}. According to \citep{Zingsem2017}, this holds,
even if it is a local minimum. In this case, the equilibrium is metastable,
such that the energy landscape presents another - stable - equilibrium.
Hence, FMR is expected to occur in metastable states. To create such
metastable states in the experiment, we sweep the angle of the applied
field at fixed field strengths. 
\begin{figure}
\includegraphics[width=1\columnwidth]{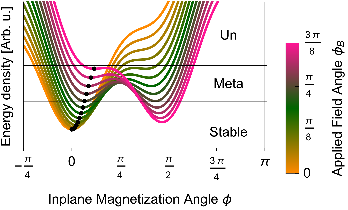}

\caption{A plot of the energy landscape at different applied field angles for
a fixed field magnitude. The black points illustrate the position
of the minimum that determines the orientation of the magnetization.
Black lines separate the unstable (marked ``Un''), metastable (marked
``Meta''), and stable (marked ``Stable'') regions for this state.\label{fig:energy-landscape-different-angles} }

\end{figure}
The effect of such an angle sweep on the energy landscape at sufficiently
low fields is illustrated in Fig. \ref{fig:energy-landscape-different-angles},
which shows the in-plane component ($\theta=\theta_{B}=\frac{\pi}{2}$)
of the Energy landscape (Eq. \ref{eq:Free-Energy-Density}) as a function
of the in-plane angle $\phi$ of the magnetization. First, a magnetic
field is applied in an easy direction at $\phi_{B}=0$ (orange curve).
Next, the direction of this applied field is changed, i.e., $\phi_{B}$
is increased. Once it is pointing parallel to the hard direction at
$\phi_{B}=\frac{\pi}{4}$ (brown curve), two states of equal energy
are present. Pushing the angle of the applied field further beyond
$\phi_{B}=\frac{\pi}{4}$ makes the state at $\phi<\frac{\pi}{4}$
unfavorable compared to the state at $\phi>\frac{\pi}{4}$. It is
now metastable. However, it is still separated from the more favorable
state at $\phi>\frac{\pi}{4}$ by an energy barrier. For even larger
angles of $\phi_{B}$, the minimum at $\phi<\frac{\pi}{4}$ turns
into a saddle point (pink curve) and becomes unstable. The black dots
mark the minimum, i.e., the magnetization's equilibrium alignment
for the state at $\phi<\frac{\pi}{4}$.

\begin{figure}
\noindent \begin{centering}
\noindent\fbox{\begin{minipage}[t]{1\columnwidth - 2\fboxsep - 2\fboxrule}%
\noindent \begin{center}
\includegraphics[width=1\columnwidth]{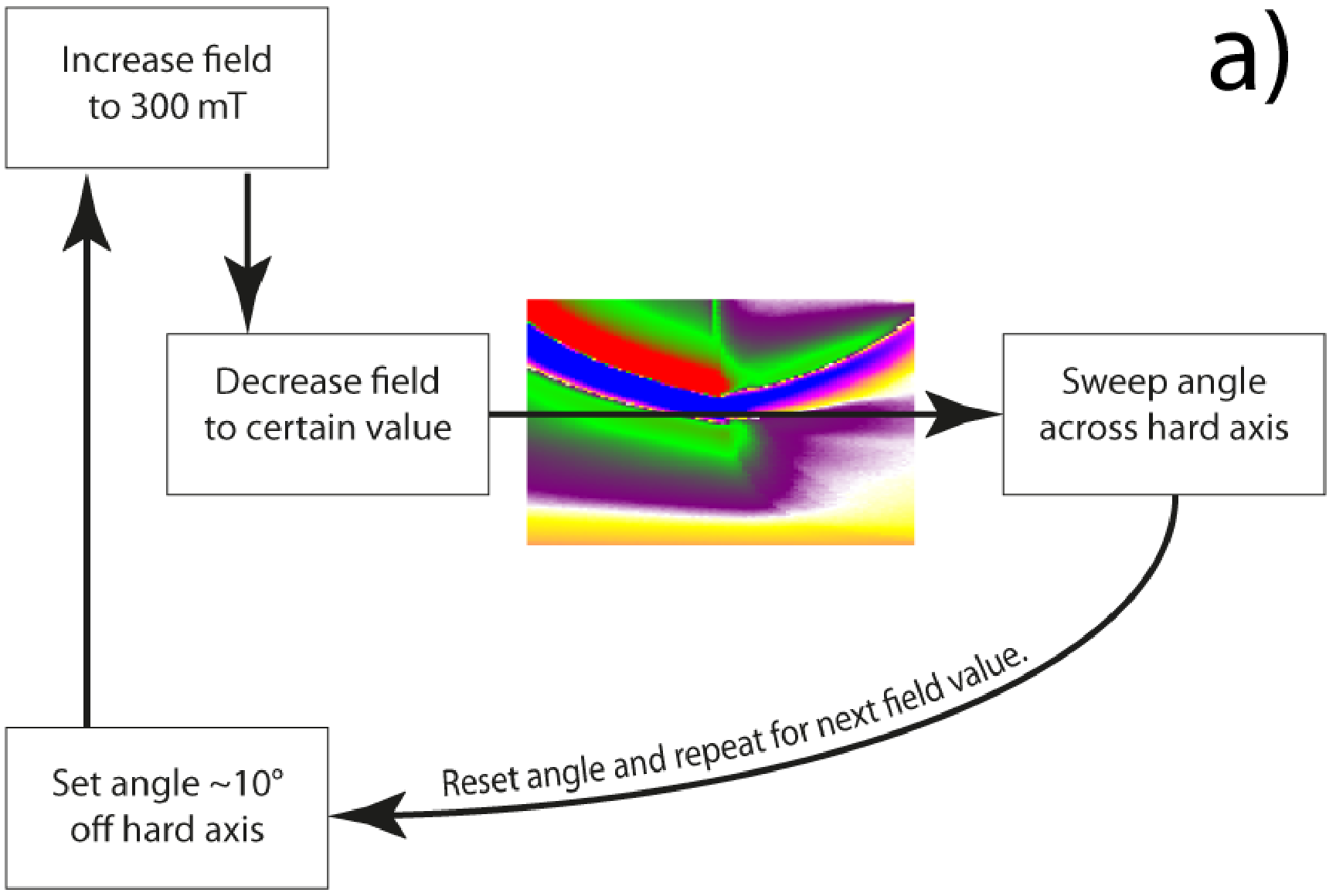}
\par\end{center}%
\end{minipage}}
\par\end{centering}
\noindent \begin{centering}
\noindent\fbox{\begin{minipage}[t]{1\columnwidth - 2\fboxsep - 2\fboxrule}%
\noindent \begin{center}
\includegraphics[width=1\columnwidth]{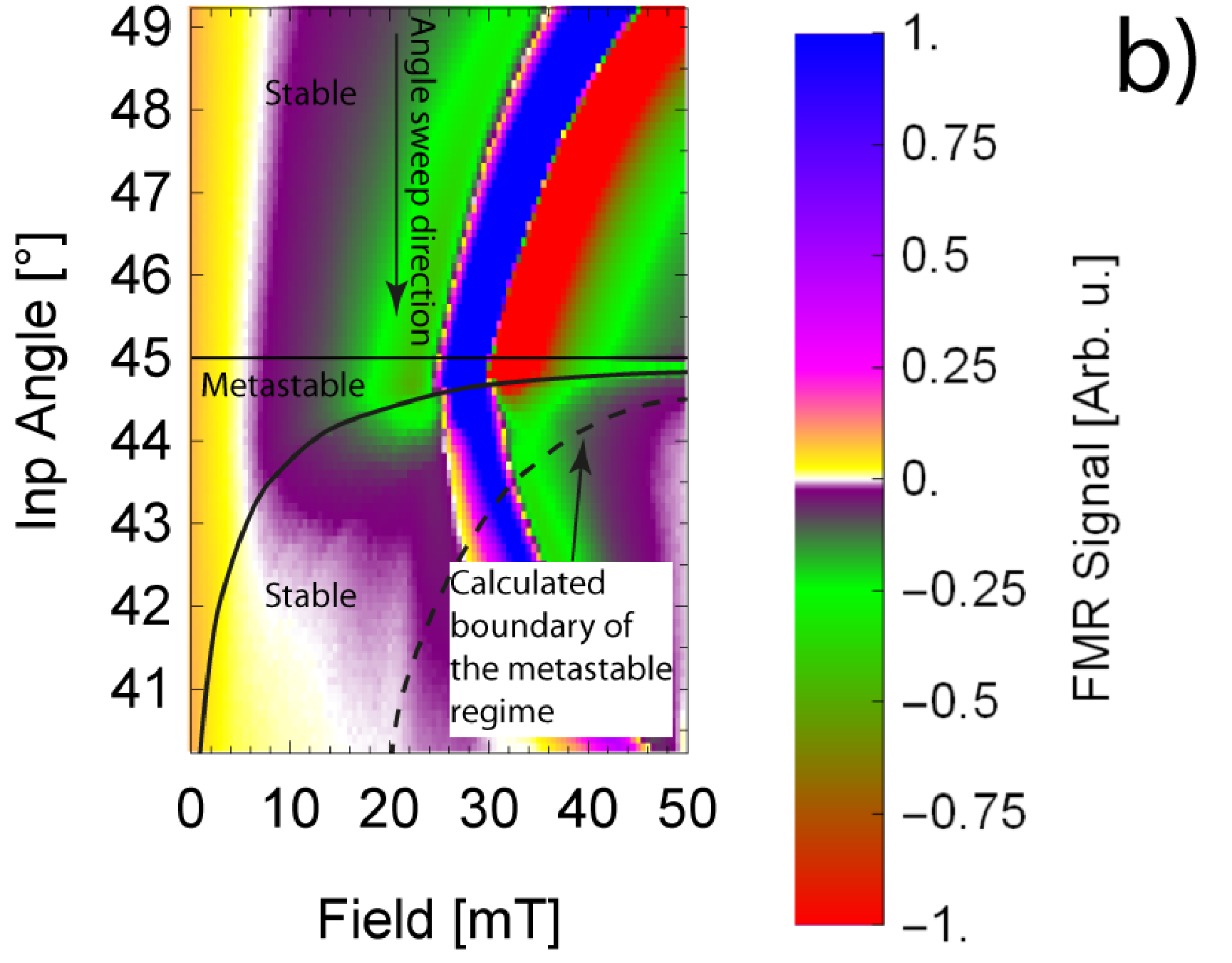}
\par\end{center}%
\end{minipage}}
\par\end{centering}
\caption{a) Schematic representation of the measurement procedure, using a
measured spectrum (compare b) ) for illustration. b) Low field section
of the measured FMR signal of a $\unit[10]{nm}$ Fe(100) film measured
at $\unit[9.535]{GHz}$ at room temperature. The straight line at
$45\text{°}$ indicates the magnetocrystalline hard direction. The
curved line shows the critical angles at which the metastable regime
ends, and the dashed curved line indicates the boundary of the metastable
regime as predicted using the theoretical model in ref. \citep{Zingsem2017}.
The difference between the black curve and the dashed curve along
the horizontal axis corresponds to the thermal fluctuation field\citep{Bance2014}.\label{fig:Schematic-representation}}
\end{figure}

The experimental procedure for such an angle sweep is schematically
shown in Fig. \ref{fig:Schematic-representation}. First, a sufficiently
large field of $\unit[300]{mT}$ is applied along the magnetocrystalline
easy axis of the sample to fully align the magnetization and overcome
all anisotropies. Then the field is reduced to the desired field value
for the angle sweep, and the angle $\phi_{B}$ is swept from the easy
axis ($\phi_{B}=0$) across the hard axis ($\phi_{B}=\frac{\pi}{4}$)
in steps of $0.1\text{°}$ to $\phi_{B}=\pi$. One angle step takes
about $\unit[0.3]{s}$ to measure. During this sweep, the FMR signal
is recorded. The sharp discontinuity in the FMR signal, which is indicated
as a black line in Fig. \ref{fig:Schematic-representation} b), pinpoints
the field-angle configurations at which the magnetization transitions
from a metastable to a stable equilibrium. For the applied field magnitudes
in this measurement, we find that the transition happens within a
few degrees off the hard axis. The dashed curve in Fig. \ref{fig:Schematic-representation}
b) shows the points at which the minimum transitions into a saddle-point,
i.e., the point at which the metastable equilibrium vanishes. The
differentce between the black curve and the dashed curve along the
horizontal axis corresponds to the thermal fluctuation field \citep{Bance2014}.
 These measurements were performed at various temperatures, and the
curves that follow the measured discontinuities at each temperature
are depicted in Fig. \ref{fig:The-boundary-of}. With increasing temperature,
the discrepancy between the points at which the transition happens,
and those at which the metastable state vanishes, becomes larger.
Hence, with increasing temperature, the magnetization overcomes larger
barriers.

\section{Results and discussion}

\begin{figure}
\noindent \begin{centering}
\includegraphics[width=1\columnwidth]{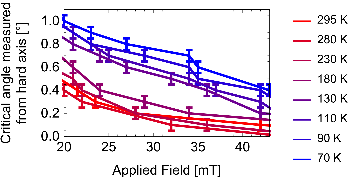}
\par\end{centering}
\caption{The boundary of the metastable regime for the sample measured in-plane
from the hard-axis of the magnetocrystalline anisotropy. Points with
error bars represent measured values. Solid lines are linear interpolations
used later to determine the temperature derivative at different fields.\label{fig:The-boundary-of}}
\end{figure}

We find that, as we decrease the temperature, the critical angle offset
from the hard axis at which the magnetization transitions from its
metastable into a stable equilibrium increases and approaches the
angle predicted for zero temperature, as shown in Fig. \ref{fig:The-boundary-of}.
The field regime from $\unit[20]{mT}$ to $\unit[43]{mT}$ was chosen
because here, the angles could be well distinguished, and the sample
is fully saturated.

\begin{figure}
\includegraphics[width=1\columnwidth]{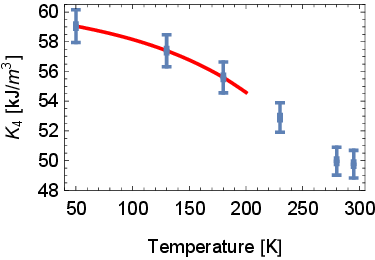}

\caption{Cubic crystal anisotropy $K_{4}$ as a function of temperature. Blue
points with error bars are the values obtained from fitting. The red
curve is a third-order polynomial interpolation to serve as a guide
to the eye.\label{fig:Fourfold-crystal-anisotropy}}
\end{figure}

For all temperatures, the resonance field was extracted from the FMR
spectra, and fitted, solving the commonly used eq. \ref{eq:FMR-Res-Cond}
\citep{Farle1998,Suhl1955,Smit1955,Giannopoulos2015} to determine
the anisotropy parameters.
\begin{equation}
\left(\frac{\omega}{\gamma}\right)^{2}=\frac{1}{M^{2}\sin^{2}\left(\theta\right)}\left(\frac{\mathrm{d}^{2}}{\mathrm{d}\theta^{2}}F\frac{\mathrm{d}^{2}}{\mathrm{d}\phi^{2}}F-\left(\frac{\mathrm{d}^{2}}{\mathrm{d}\theta\mathrm{d}\phi}F\right)^{2}\right)\label{eq:FMR-Res-Cond}
\end{equation}
 For the fits, the experimental resonance field was used in the calculation
to determine the orientation of the magnetization vector that minimizes
the free energy density. This magnetization vector was then used in
Eq. \ref{eq:FMR-Res-Cond} to calculate the resonance field position
for the respective frequency at a given set of anisotropy parameters.
The so determined cubic anisotropy shown in Fig. \ref{fig:Fourfold-crystal-anisotropy}
is in agreement with literature data \citep{Farle1998}. 
 The out of plane uniaxial anisotropy $K_{2}^{\bot}$ and the magnetization
did not change significantly in the given temperature regime. We obtain
$M=\unitfrac[(1.71\pm0.01)\cdot10^{6}]{A}{m}$ and $K_{2}^{\bot}=\unitfrac[(16\pm0.6)\cdot10^{3}]{J}{m^{3}}$,
using a g-factor of $2.09$\citep{Frait1971}.

\begin{figure}
\includegraphics[width=1\columnwidth]{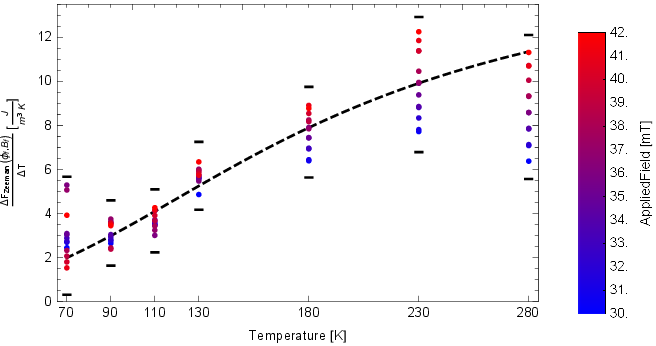}

\caption{Magnon specific heat as a function of temperature. The red points
show the central numerical derivative of the Zeeman contribution to
the free energy including error bars. The black dashed curve is the
result of numerically evaluating the magnon specific heat eq. \ref{eq:specific-heat-magnons}.
The parameters we used for Fe are $a=\unit[286.65]{pm}$ \citep{latticFe}
$\omega_{ZB}=\unit[18]{THz}$ \citep{Kittel1963} $\gamma=\unit[2.92\cdot10^{10}]{\nicefrac{1}{T\cdot s}}$
\citep{Frait1971} and $\alpha_{D}=0.87$ \citep{Rezende2014a}.\label{fig:Zeeman-temp-deriv}}
\end{figure}

With these values, we then evaluated the contribution of the Zeeman-energy
$F_{\mathrm{Zeeman}}=-\vec{M}\cdot\vec{B}$ to the free energy density
in Eq. \ref{eq:Free-Energy-Density} for all temperatures using the
critical angles $\phi_{B}$ from Fig. \ref{fig:The-boundary-of} as
the in-plane applied field angle. In this calculation, we fixed $\theta_{B}$
and $\theta$ to $90\text{°}$ as they are constrained by the sample's
shape anisotropy. The magnetization angle $\phi$ is determined numerically
by minimization eq. \ref{eq:Free-Energy-Density} within the metastable
regime. This critical Zeeman energy density is the energy density
that is provided to the system in order to perform the transition.
Therefore we analyze its change as a function of temperature. Considering
that the magnetization has some amount of heat available to it in
the form of magnons, we would expect that, as we decrease the temperature,
more Zeeman energy is required to make the transition. Moreover, since
the Zeeman contribution is defined negative, we expect that its change
is proportional to the change of the magnon heat in the system. Therefore
we calculated the change of the thermal energy of the magnons that
is the heat capacity. To calculate the magnonic heat capacity, we
proceed as described in \citep{Kittel1963}. As of \citep{Rezende2014}
the magnon dispersion can be approximated as 
\begin{equation}
\omega\left(k\right)=\gamma B+\omega_{ZB}(1-\cos\left(\frac{\pi}{2}\frac{k}{k_{m}}\right))\label{eq:magnon-dispersion}
\end{equation}
where $\gamma$ is the magnetogyric ratio, $B$ is the magnetic flux,
$k_{m}=\alpha_{D}\frac{\sqrt[3]{6\pi^{2}}}{a}$ is the radius of the
Debye-Sphere with the lattice constant $a$ and the scaling factor
$\alpha_{D}$ \citep{Rezende2014a} that approximates the Brillouin
zone, and $\omega_{ZB}$ is the magnon frequency at the zone boundary.
According to \citep{Kittel1963}, the magnon specific heat can then
be written as 
\begin{equation}
c_{V}^{\mathrm{magnon}}=\frac{1}{\left(2\pi\right)^{3}}\intop_{0}^{k_{m}}\mathrm{d^{3}}k\frac{\left(\hbar\omega\left(k\right)\right)^{2}}{k_{b}T^{2}}\frac{\exp\left(\frac{\hbar\omega\left(k\right)}{k_{B}T}\right)}{\left(\exp\left(\frac{\hbar\omega\left(k\right)}{k_{B}T}\right)-1\right)^{2}}\label{eq:specific-heat-magnons}
\end{equation}
by calculating the temperature derivative of the inner energy of the
magnons. A comparison between the temperature derivative of the Zeeman
contribution and the numerically calculated magnon heat capacity of
Fe is shown in Fig. \ref{fig:Zeeman-temp-deriv}. We find that, for
each applied field, the curves are proportional and the data is in
line with the models in \citep{Wojtczak1983} and \citep{Kittel1963}.
This leads us to conclude that this experiment grants direct access
to the heat capacity of magnons at elevated temperatures.
 This technique can be applied to any ferromagnet since any ferromagnet
can be tailored to have any desired anisotropy simply by shape, even
if the intrinsic magnetocrystalline anisotropy is small.

\section{Summary}

In a magnetization configuration non-collinear with the external field,
we have shown that FMR modes exist in metastable magnetic states as
predicted in \citep{Zingsem2017}. We also find that the magnonic
heat capacity of iron is proportional to the temperature derivative
of the Zeeman energy at the critical points in the unconventional
FMR angular dependence by comparison to spin-wave theory calculations.
We find a good agreement between the measured data and the calculation.
The magnitude of of the magnon contribution to the specific heat shown
in Fig. \ref{fig:Zeeman-temp-deriv} is also in line with other works\citep{Rezende2014,Boona2014,Rezende2014a,Rezende2015}.
Our results suggest that measuring the temperature-dependent size
of the energy barrier, which confines a saturated meta-stable magnetic
state, presents a new method for determining the magnon contribution
to the specific heat. The only requirement for these measurements
is that the sample exhibits magnetic anisotropy. This can either be
magnetocrystalline anisotropy or shape anisotropy in patterned magnetic
shapes.
\begin{acknowledgments}
In part funded by the Deutsche Forschungsgemeinschaft (DFG, German
Research Foundation) -- Project-ID 405553726 -- TRR 270\textquotedblleft .
This study was supported in part by the Research Grant No. 075-15-2019-1886
from the Government of the Russian Federation.
\end{acknowledgments}

\bibliographystyle{IEEEtran}
\bibliography{literature}

\end{document}